\documentclass[pra,twocolumn,showpacs,superscriptaddress,amsmath,amssymb]{revtex4}
\usepackage{graphicx}

\usepackage{color}
\begin{document}
\title{Exact results on the Kitaev model on a hexagonal lattice:
spin states, string and brane correlators, and anyonic excitations}
\author{Han-Dong Chen}
\affiliation{Department of Physics, University of Illinois at Urbana-Champaign, Urbana, IL 61801}
\author{Zohar Nussinov}
\affiliation{Department of Physics, Washington University, St.
Louis, MO 63160, USA}
\date{\today}
\begin{abstract}
In this work, we illustrate how a Jordan-Wigner
transformation combined with symmetry considerations
enables a direct solution of Kitaev's model on 
the honeycomb lattice. We (i) express the 
p-wave type fermionic ground states of
this system in terms of the original
spins, (ii) adduce that symmetry alone 
dictates the existence of 
string and planar {\em brane}
type correlators and their
composites, (iii) compute
the value of such non-local correlators
by employing the Jordan-
Wigner transformation, (iv) affirm
that the spectrum is inconsequential
to the existence of topological quantum
order and that such information is encoded
in the states themselves, and (v) express the 
the local symmetries of Kitaev's model and the anyonic character
of the excitations in terms of fermions.

\end{abstract}

\maketitle

\section{Introduction}

Topological quantum order (TQO) \cite{wenbook}
\cite{Kitaev2003} is a new paradigm that
lies beyond the realm of Landau's theory 
\cite{Landau} \cite{andersonrev} of 
local order parameters.  TQO is intuitively
associated with insensitivity to {\it local} perturbations: the order
is {\it topological}.  As such, TQO cannot be described by {\it local}
order parameters. A quintessential example
of a system with TQO is Kitaev's model on the honeycomb lattice
\cite{Kitaev2006} \cite{Pachos} \cite{Vidal}. In this article, we will study
this model. Our {\em new central results} are (a) an 
explicit form for the ground states in real space
which extends and complements the works of 
\cite{Kitaev2006} and \cite{Pachos} as well as (b) our finding of 
non-local correlations [two dimensional 
{\em string} or {\em brane} type correlators] 
in this system. We further show how the (c) 
known anyonic excitations of Kitaev's model 
can be examined anew by using a direct Jordan-Wigner
transformation. Beyond providing a direct solution which 
highlights certain previously overlooked 
aspects, our results will flesh out some of the more general ideas 
\cite{NO}, \cite{Chen2007} regarding the 
general character of TQO.

\section{Outline}

This article is organized as follows: In section \ref{section-fermionization}, we review the fermionization of the Kitaev model on a hexagonal lattice\cite{Feng2006,Chen2007}. The original model of spins on a hexagonal lattice is mapped to a model of $p$-wave BCS model with site-dependent chemical potential for spinless fermions on a square lattice. In section \ref{section-symmetry}, we discuss the fermionic representation of symmetries embodied in the model. It is shown that the local conserved quantities on the plaquettes studied by Kitaev\cite{Kitaev2006} are equivalent to the conserved  bond quantities in the fermionic representation, up to a gauge fixing. In section \ref{section-groundstate}, the ground state configuration, which is vortex free in spin representation and with a uniform chemical potential in fermionic representation, is exactly solved in fermionic representation. The spin basis form of the ground state is also studied. It is shown that the ground state can be written as a projection of a given reference state over a sector of Hilbert space that has a chosen set of topological numbers which are similar to, yet slightly more complicated than, 
other systems which exhibit TQO such as the Rokhsar-Kivelson dimer model\cite{RK} and the Kitaev toric code model\cite{Kitaev2003}. The phase diagram and Bogliubov excitations are also obtained. In Section(\ref{section-anyon}), we show
how the known results about the anyonic vortex excitation in gapped state can be easily studied in the fermionic representation. In section \ref{section-correlators}, a symmetry argument is put forward concerning the vanishing correlation functions, which is also recently derived anew by a Majorana fermionization construction\cite{bms}. We also show that string correlators naturally appear in Kitaev's model.A brief summary in section \ref{section-conclusion} concludes this work. In the appendices, we provide technical details concerning the 
determination of the ground state and review Elitzur's theorem.

\section{Fermionization}\label{section-fermionization}

\begin{figure}[h]
\includegraphics[width=2.5in]{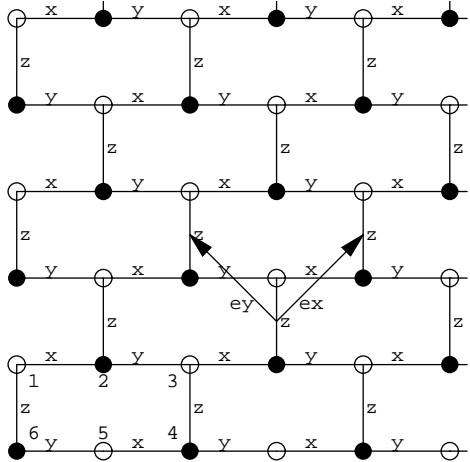}
\caption{Deformed hexagonal lattice and three types of bonds.}
\label{FIG-hexagonal}
\end{figure}

We begin with a Fermionization of the Kitaev model on the 
hexagonal lattice \cite{Kitaev2006} \cite{Pachos} which is 
defined by the following $S=1/2$ Hamiltonian
\begin{eqnarray}
H&=&-J_x\sum_{x-bonds} \sigma^x_{R}\sigma^x_{R'}
-J_y\sum_{y-bonds} \sigma^y_{R}\sigma^y_{R'}\nonumber\\
&&-J_z\sum_{z-bonds} \sigma^z_{R}\sigma^z_{R'},\label{H}
\end{eqnarray}
with $R$ and $R'$ lattice sites. 
This system can be 
fermionized by a
Jordan-Wigner transformation\cite{Feng2006,Chen2007}.  
This one-dimensional fermionization is made 
vivid by deforming the hexagonal lattice 
into a ``brick-wall lattice'' which is topologically
equivalent to it on which 
we may perform a one dimensional Jordan-Wigner transformation.
The schematics are shown in Fig.\ref{FIG-hexagonal}. 
In the up and coming, we will mark all sites 
by ``white'' or ``black'' (w/b)in order to denote to which 
sublattice which they belong to.
The distance between two nearest-neighboring 
sites on this lattice will be set to unity.

\begin{figure}
\includegraphics[width=2.5in]{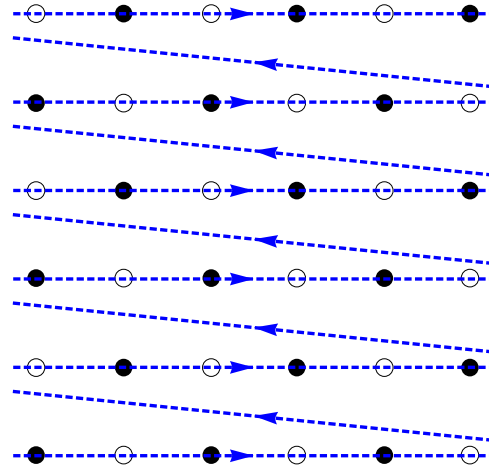}
\caption{Schematics of the contour for the
Jordan-Wigner transformation that we employ
in the deformed hexagonal lattice- see text and Eq.(\ref{JWc})
in particular.}
\label{FIG-contour}
\end{figure}

Throughout, we will consider the system
with open boundary conditions unless stated otherwise.  
The diagonal directions $\hat{e}_x$ 
and $\hat{e}_y$  shown in Fig.(\ref{FIG-hexagonal}) will be of 
paramount importance in our final solution.
Let us denote by $(i,j)$ the Cartesian coordinates of each 
site $R$ on the lattice of Fig. \ref{FIG-hexagonal}. Let us next consider
the Jordan-Wigner transformation defined by a simple one dimensional contour
which threads the entire lattice [See Fig.\ref{FIG-contour}]:
\begin{eqnarray}
\sigma^+_{ij}&=&2\left[\prod_{j'<j}\prod_{i'}\sigma^z_{i'j'}\right]
\left[\prod_{i'<i}\sigma^z_{i'j}\right]c^\dag_{ij} \nonumber
\\ \sigma^z_{ij}&=&2c^\dag_{ij}c^{}_{ij}-1.
\label{JWc}
\end{eqnarray}
This path goes through each 
lattice site exactly once
as shown in Fig.(\ref{FIG-contour}).
In Eq.(\ref{JWc}), $\sigma^{+} = (\sigma_{x} 
+ i \sigma^{y})$ is twice the spin raising operator at a given site-
hence the factor of two.

The Kitaev model of Eq.(\ref{H}) now becomes
\begin{eqnarray}
H&=&J_x\sum_{x-bonds} 
\left(c^\dag-c^{}\right)_w\left(c^\dag+c^{}\right)_b\nonumber\\
&&-J_y\sum_{y-bonds} \left(c^\dag+c^{}\right)_b\left(c^\dag-c^{}\right)_w\nonumber\\
&&-J_z\sum_{z-bonds} 
\left(2c^\dag c^{}-1\right)_b\left(2c^\dag c^{}-1\right)_w.
\end{eqnarray}
Henceforth, the subscripts $b$ and $w$ will denote 
the white and black sites of a bond as 
illustrated in Fig.\ref{FIG-hexagonal}.

Let us next introduce the Majorana fermions
\begin{eqnarray}
A_w=(c-c^\dag)_w/i \quad B_w=(c+c^\dag)_w
\end{eqnarray}
for the white sites and
\begin{eqnarray}
B_b=(c-c^\dag)_b/i\quad A_b=(c+c^\dag)_b
\end{eqnarray}
for the black sites.
With all of these transformations in tow,
the Hamiltonian now reads
\begin{eqnarray}
H&=&-i\left[\sum_{x-bonds} J_x A_w^{}A_b^{} - \sum_{y-bonds}J_y A_b^{}A_w^{}\right]
\nonumber\\
&&-J_z\sum_{z-bonds} J_z (BA)_b(BA)_w.
\label{lm}
\end{eqnarray}
It is easy to see that $BB$ along the $z$-bond is a conserved quantity
\cite{Feng2006}. Thus, the $Z_{2}$ operator 
\begin{eqnarray}
\alpha_{r}=iB_{b} B_{w},
\label{aBB}
\end{eqnarray} 
with $r$ the coordinate of the midpoint of
the bond connecting the black and red sites,
is fixed for each vertical bond. 
The Hamiltonian of Eq.(\ref{H}) now reads
\begin{eqnarray}
H(\{\alpha\})&=&-i\left[\sum_{x-bonds} J_x A_w^{}A_b^{} - \sum_{y-bonds}J_y A_b^{}A_w^{}\right]
\nonumber\\
&&-iJ_z\sum_{z-bonds}  \alpha_{r} A_b A_w.
\end{eqnarray}

Here, $r$ denote the centers of the 
vertical bonds. In Section(\ref{section-symmetry}),
We will show that 
$\{\alpha_r\}$ are intimately related to the local symmetries
present in Kitaev's model of Eq.(\ref{H}).
This identification, combined with Reflection Positivity
arguments, \cite{Kitaev2006} will allow us to infer that,
up to ($d=1$ \cite{BN}) symmetry operations,  
$\alpha_{r} =1$ for all r. The ground state
does not contain any ``vortices'' 
which are marked by one dimensional 
in the Ising variables $\{\alpha_{r}\}$ along 
a row. Similar Reflection Positivity
arguments regarding the absence of vortices in 
other systems and a bound
on the energy penalties that they entail
are. e.g., given in \cite{XY}. 
This, in turn, will allow us to explicitly
diagonalize the Hamiltonian.

\section{Fermionic representation of local Symmetries}
\label{section-symmetry}
As shown by Kitaev\cite{Kitaev2006}, the Hamiltonian of
Eq.(\ref{H}) has one conserved quantity for each plaquette (or hexagon) $h$, 
\begin{eqnarray}
I_{h}=\sigma^y_{1w}\sigma^z_{2b}\sigma^x_{3w}\sigma^y_{4b}\sigma^z_{5w}\sigma^x_{6b}.
\label{Eq-kitaev-conserved}
\end{eqnarray}
Here, $1-6$ denote the sites of a given plaquette, as illustrated in the plaquette of the lower corner in Fig.\ref{FIG-hexagonal}. In the subscripts, 
we also label the (w/b) sublattices of the sites. 
The conserved quantity $I_{h}$ on plaquette is equivalent to the bond conserved quantity $\alpha$ defined in previous section, up to a gauge fixing as we shall show below. This can be shown by fermionizing $I_{h}$ using the transformation we introduced in previous section. Let us first fermionize the product of first three spins
\begin{eqnarray}
\sigma^y_{1w}\sigma^z_{2b}\sigma^x_{3w}&=&\frac{1}{i}\left(c^\dag-c^{}\right)_{1w}
\sigma^z_{2b}\sigma_{1w}^z\sigma_{2b}^z\left(c^\dag+c^{}\right)_{3w}
\nonumber\\
&=&i\left(c^\dag+c^{}\right)_{1w}
\left(c^\dag+c^{}\right)_{3w}\nonumber\\
&=&iB_{1w}B_{3w}.
\end{eqnarray}
Similarly,
\begin{eqnarray}
\sigma^x_{6b}\sigma^z_{5w}\sigma^y_{4b}=iB_{4b}B_{6b}.
\end{eqnarray}
Therefore, 
\begin{eqnarray}
I_{h}=\alpha_{34}\alpha_{16}.
\label{Ihalpha}
\end{eqnarray}

This model has an extra degeneracy that links difference sectors parameterized by different sets of $\alpha$'s. The vortex [or anyon] variables
are the product of two consecutive Ising bond variables.
In other words, anyons ($I_{h}$) are none other than domain walls
in the Ising variables ($\{\alpha_{r}\}$) that our system contains. 
Inverting all of the values of $\alpha_{r}$ for
all sites $r$ which lie along a horizontal row leaves the 
system unchanged. Physically, effecting 
the transformation 
$\alpha_{r}  \to - \alpha_{r}$ for all bonds $r$ along
a row does not change the vorticity content of the 
system: all domain walls along the chain remain invariant
(and as we show so does the spectrum). In the notation of \cite{NO} and 
\cite{BN}, this corresponds to a $d=1$ dimensional operator [as
it involves spin operators on a ($d=1$ dimensional) line]. 
 For simplicity, let us divide  $\alpha$ into different subsets: $\{\alpha\}=\cup_i[\alpha]_i$. Here, $[\alpha]_i$ denotes the set of bonds that are connected to white sites of the $i$-the horizontal line of the brick-wall lattice. 
We next explicitly write down these $d=1$ symmetry operators
(see \cite{NO}, \cite{BN} for a definition of $d$ dimensional
symmetry operators)
in both their fermionic and original spin
language form.  Towards this end, we construct the unitary operator $U_i^w$
\begin{eqnarray}
U_i^w = \prod_{j\geq i}\prod_{n \in  j}A_{n}.
\label{uiw}
\end{eqnarray}
In Eq.(\ref{uiw}), $n$ is a site index. The product
is taken over all sites in the rows $j \geq i$.  
We notice that $U_i^w$ effectively reverses the sign of $[\alpha]_i$ while leaving others untouched
\begin{eqnarray}
\left(U_i^w\right)^\dag H(...,[\alpha]_{i-1},[\alpha]_i,[\alpha]_{i+1},...)\left(U_i^w \right)\nonumber\\
= H(...,[\alpha]_{i-1},-[\alpha]_i,[\alpha]_{i+1},...).
\end{eqnarray}
Therefore, we have extra freedom to fix one $\alpha$ in each subset $[\alpha]_i$. This degree of freedom is closely related to the string-like 
conserved quantity in the original spin model, namely, 
\begin{eqnarray}
P_j=\prod_i \sigma^z_{ij},
\label{pz}
\end{eqnarray}
which rotates all spins on $j$-th line by 180 degree around $z$.
Under the gauge fix in which one $\alpha$ is fixed to $+1$ in each $[\alpha]_i$,  the set of Kitaev conserved quantities 
$\{I_{h}\}$ is equivalent to $\{\alpha\}$. 

\begin{figure}
\includegraphics[width=2.8in]{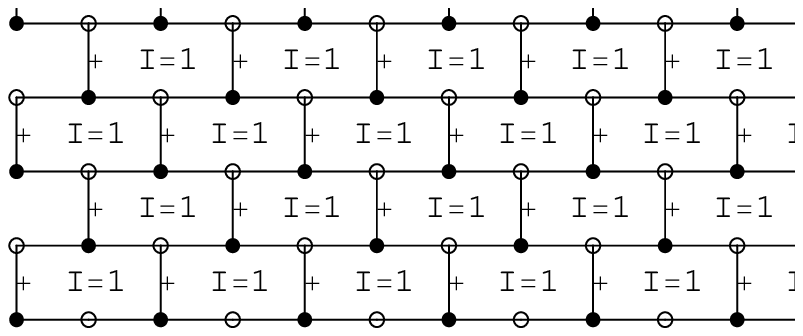}\\ (a) \\ ~ \\
\includegraphics[width=2.8in]{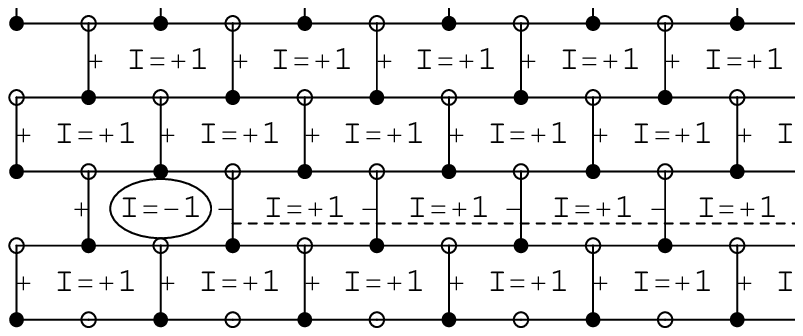}\\ (b) 
\caption{(a) Vortex free configuration. (b)Reversing $I$ in a plaquette corresponds to flip the chemical potential along a string.}\label{FIG-states}
\end{figure}

\section{Ground state}\label{section-groundstate}

\subsection{Diagonalization}

Armed with a physical meaning of the bond variables $\alpha$
from Section(\ref{section-symmetry}), we now proceed to solve
the problem posed by Eq.(\ref{lm}) and diagonalize the problem.

It is interesting to introduce a fermion on the $z$-bond 
\begin{eqnarray}
d=(A_w+i A_b)/2, \quad d^\dag = (A_w-iA_b)/2,
\end{eqnarray}
where $A_w$ and $A_b$ are the Majorana fermions on the white and black site of a given $z$-bond. We thus have a model for fermions on a square lattice with 
a site dependent chemical potential
\begin{eqnarray}
H &=&  J_x\sum_{r}\left(d_r^\dag + d_r^{}\right)
\left(d_{r+\hat{e}_x}^\dag - d_{r+\hat{e}_x}^{}\right)\nonumber\\
&&+J_y\sum_{r}\left(d_r^\dag + d_r^{}\right)
\left(d_{r+\hat{e}_y}^\dag - d_{r+\hat{e}_y}^{}\right)
\nonumber\\
&&+J_z\sum_r \alpha_r (2d^\dag_r d^{}_r-1).\label{EQ-fermion-model}
\end{eqnarray}
The unit vector
$\hat{e}_y$ connects two $z$-bonds and crosses a $y$-bond,see 
Fig.(\ref{FIG-hexagonal}). A similar definition holds for $\hat{e}_x$. 
Note that 
\begin{eqnarray}
[\alpha_{r}, d_{r}] = [\alpha_{r}, d^{\dagger}_{r}]=0.
\end{eqnarray}
For sufficiently large systems,
the ground state configurations are bulk vortex-free 
configurations \cite{Kitaev2006} in which $I_{h}=1$ for all plaquettes 
(hexagons) $h$. The ground states of the fermionic problem
of Eq.(\ref{EQ-fermion-model})
has $\alpha_{r}=1$ everwhere [and all
sectors of $\{\alpha_{r}\}$ related to
it by the $d=1$ operation of Eqs.(\ref{uiw},\ref{pz})]
which corresponds
to $I_{h}=1$ for all plaquettes 
$h$. As shown in Fig.(\ref{FIG-states}), reversing $I_{h}$ to
$I_{h}=-1$ leads to an inversion of the chemical
potentials $\{\alpha_{r}\}$ along a horizontal string. 

The exact solution for ground state is now easy to obtain for the bulk 
system by a Fourier transformation. Up to innocuous additive constants, 
the Hamiltonian of the vortex-free configuration now reads, 
\begin{eqnarray}
H_g=\sum_{q}\left[\epsilon_q d^\dag_q d^{}_q + i\frac{\Delta_q}{2}\left(d^\dag_q d^{\dag}_{-q}+H.c.\right)\right],
\end{eqnarray}
where \begin{eqnarray} 
\epsilon_q&=&2J_z-2J_x\cos q_x-2J_y\cos q_y, \nonumber
\\ \Delta_q&=&2J_x\sin q_x+2J_y\sin q_y.
\label{eD}
\end{eqnarray}
The fermionized Hamiltonian (\ref{EQ-fermion-model}) describes a 
$p$-wave type BCS pairing model with site-dependent chemical potential.
After a Bogliubov transformation, this Hamiltonian can be diagonalized 
and the quasiparticle excitation is 
\begin{eqnarray}
E_q=\sqrt{\epsilon_q^2+\Delta_q^2}.\label{ks}
\end{eqnarray}

\subsection{Spectrum}

The energy spectrum, in the low energy 
vortex-less sector $\alpha_r=1$ [and all
sectors of $\{\alpha_{r}\}$ related to
it by the $d=1$ operation of Eqs.(\ref{uiw},\ref{pz})], which we found 
to be given by Eq.(\ref{eD}, \ref{ks}) obviously does not encode
information about the topological nature of Kitaev's model. 
The equivalence of the lowest eigenvalues of the Hamiltonians in both 
{\bf (i)} a simple BCS type problem given by Eqs.(\ref{eD},\ref{ks})
and {\bf (ii)} the Kitaev model of Eq.(\ref{H}) vividly illustrates 
the maxim that the states themselves in a particular (operator language) representation and not their energies which 
determine whether or not TQO exists. \cite{NO}, \cite{Chen2007}
In other words, the same Hamiltonian in different 
representations which are related to one another by unitary transformations 
and thus preserve the same set of eigenvalues (the energy spectrum) \cite{NO} 
can describe both topologically quantum ordered systems 
(such as Kitaev's model) or systems with no topological order.\cite{NO} 
\cite{Chen2007} In the current context, these mappings
are the Jordan-Wigner transformations that we apply. \cite{Chen2007}

\begin{figure}
\includegraphics[width=2.8in]{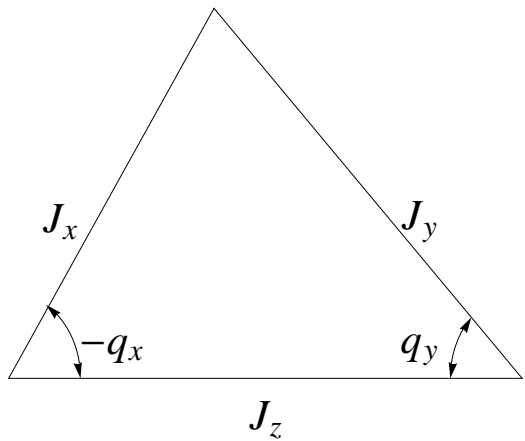}\\ 
\caption{A geometric interpretation of Eqs.(\ref{eD})
for the existence of gapless excitations. The 
two conditions $\epsilon_{q} = \Delta_{q}=0$ 
imply that $q_{x,y}$ can be regarded
as angles in the triangle formed by $\{J_{x},J_{y},J_{z}\}$ as shown. 
The law of cosines gives Eqs.(\ref{coeq}). This plot immediately
leads to the triangle inequality condition of Eq.(\ref{tin}).
When the triangle inequality is violated (the gapped 
phase), $q_{x,y}$ are imaginary and lead to a finite
correlation length as we will discuss later
[Eq.(\ref{corli})].}\label{FIG-triangle}
\end{figure}

Let us now proceed to study when the spectrum of Eqs.(\ref{eD}, \ref{ks}) describes
a system with a spectral gap between the ground and 
the next excited states and delineate these
boundaries between the gapped and gapless phases.
Towards this end- see Fig.(\ref{FIG-triangle})- 
we study the nominal solution to $E_q=0$ which is 
\begin{eqnarray}
q_x&=&\pm\arccos\left[\frac{J_x^2+J_z^2-J_y^2}{2J_x J_z}\right],  \nonumber
\\ q_y&=&\pm\arccos\left[\frac{ J_y^2+ J_z^2-J_{x}^{2}}
{2J_y J_z}\right].\label{coeq}
\end{eqnarray}

This solution makes sense only when the conditions
\begin{eqnarray}
&&-2|J_y| |J_z|\leq J_y^2+J_z^2-J_x^2 \leq 2|J_y| |J_z|,\\
&&-2|J_x| |J_z|\leq J_x^2+J_z^2-J_y^2 \leq 2|J_x| |J_z|
\end{eqnarray}
are satisfied simultaneously. These inequalities can be easily 
solved to obtain the condition for gapless excitations
\begin{eqnarray}
&&|J_x|\leq |J_y|+|J_z|,\\
&&|J_y|\leq |J_x|+|J_z|,\\
&&|J_z|\leq |J_x|+|J_z|.
\label{tin}
\end{eqnarray}
This result is in agreement with Kitaev's original 
solution and Pachos' analysis. \cite{Kitaev2006} \cite{Pachos}
We can now view $q_{x,y}$ as (up to an inversion $q_{a} \to - q_{a}$ for one
of the components of $\vec{q}$)
as the angles in the triangle formed by the couplings
$J_{x}, J_{y}$ and $J_{z}$- see Fig.(\ref{FIG-triangle}). 
Eq.(\ref{coeq}) is the law
of cosines in this triangle which automatically satisfies $\Delta_{q} 
= \epsilon_{q} =0$ of Eq.(\ref{eD}) and consequently has 
$E_{q} =0$. It is also interesting 
to notice that the cyclic symmetry between $J_x, J_y, J_z$ 
is explicitly restored in the ground state solution although 
its explicitness  is lost in the fermionized model.

Next, we briefly remark on not only the low energy spectrum of 
pertinence to the zero temperature problem, but
rather examine the entire spectrum of
the theory. After tracing over the fermionic
degrees of freedom present in Eq.(\ref{EQ-fermion-model}),
we obtain an effective two dimensional 
Ising type Hamiltonian (in $\{\alpha_{r}\}$) 
with long range interactions. The full spectrum
of this long range Ising type Hamiltonian is 
identical to that of Kitaev's model.

\subsection{Real space form of the ground states}

 The BCS type ground state corresponding to Eq.(\ref{ks}) is
\begin{eqnarray}
| g \rangle = \prod_{k} \left( u_{k} + v_{k} d_{k}^{\dagger} d_{-k}^{\dagger} \right) | 0 \rangle,
\label{soln2kit2}
\end{eqnarray}
where 
\begin{eqnarray}
|v_{k}|^{2} = \frac{1}{2} \left[ 1 - \frac{\epsilon_{k}}{E_{k}} \right], 
\quad\quad |u_{k}|^{2} = \frac{1}{2} \left[ 1 + \frac{\epsilon_{k}}{E_{k}} \right].
\label{uv}
\end{eqnarray}
Let us now invert the transformations that we have performed
until now (Jordan-Wigner and others) in order to express
the Fermionic operators and the Fermionic vacuum state 
in terms of the original spin degrees of freedom. 
In what follows, we will, when needed, keep explicit track
of the (w/b) sublattices of each of the sites of 
each vertical bond whose center is at
$\vec{r}$. Undoing all of the transformations that we employed
thus far, we
have
\begin{eqnarray}
d_{k}^{\dagger} = - \frac{1}{2} \sum_{r_{w}} \left[ \sigma^{y}_{r_{w}}
\left( \prod_{r'<r_{w}} \sigma_{r'}^{z} \right) - i \sigma^{x}_{r_{b}}
\left( \prod_{r'<r_{b}} \sigma^{z}_{r'} \right) \right] \nonumber
\\ \times e^{- i \vec{k} \cdot
(\vec{r}_{w} - \frac{1}{2} \hat{e}_{z})}.
\label{dkf}
\end{eqnarray}
[For a definition of $\hat{e}_{z}$ see Fig.(\ref{FIG-hexagonal}).]
The product $\prod_{r'<r_{w}}$ corresponds
to the product of all lattice sites $\vec{r}'$ which appear
before $r_{w}$ on the Jordan-Wigner contour 
of Eq.(\ref{JWc}) which traverses all sites of the two dimensional
lattice.A similar definition
applies to $\prod_{r'<r_{b}}$: it is the product over
all lattice sites which appear before $r_{b}$
in the Jordan-Wigner product. 

In the spin basis, a fermionic vacuum corresponds to 
\begin{eqnarray}
|0 \rangle = {\cal{N}} \left( \prod_{r_{w}} \left[ \frac{1}{2} 
(1+ \mathcal{B}_{r_{w}}) \right] \right) \left( \prod_{h} 
\left[ \frac{1}{2} (1+ I_h) \right] \right) \nonumber
\\ \times
\Big(
\prod_{j} \left[ \frac{1}{2} (1+ \overline{\alpha}_{r^{*}_{jw}} ) \right] 
\Big) 
| \phi \rangle.
\label{fermion_vac}
\end{eqnarray}  
In Eq.(\ref{fermion_vac}), the product over $h$ is that 
over all elementary hexagons, $|\phi \rangle$ is an arbitrary
reference state: e.g. in the $\sigma^{z}$ basis, we may choose
it to be the fully polarized state $|\phi \rangle = | \uparrow
\uparrow ... \uparrow \rangle$, the operator
\begin{eqnarray}
\mathcal{B}_{r_{w}} \equiv - \sigma^{x}_{r_{w}} \sigma^{z}_{r_{w}+1} ... \sigma^{z}_{r_{b}-1} \sigma^{x}_{r_{b}}
\label{brw}
\end{eqnarray}
extends over all sites lying between (and including) $r_{w}$ and $r_{b}$ as 
labeled by the one dimensional Jordan-Wigner contour that connects $r_{w}$ with $r_{b}$, ${\cal{N}}$ is a normalization factor,
and the hexagonal operator $I_h$ as given by Eq.(\ref{Eq-kitaev-conserved}). 
Similarly, inverting the fermionization carried earlier,
we find that
\begin{eqnarray}
\overline{\alpha}_{r_{w}} = \Big( \prod_{l <r_{w}} \sigma^{z}_{l} \Big) 
\Big[ -\sigma_{r_{w}}^{x} + \Big( \prod_{r_{w} \le l^{\prime} < r_{b}} 
\sigma_{l^{\prime}}^{z} \Big) (- i \sigma_{r_{b}}^{y}) \Big]
\label{alphaspin}
\end{eqnarray}
is the spin representation of the operator $\alpha_{r}$ of Eq.(\ref{aBB}).
Eq.(\ref{alphaspin}) defines an operator $\overline{\alpha}_{r_{w}}$ for
any white lattice site $r_{w}$. In the last product in Eq.(\ref{fermion_vac}),
we have a product of $\frac{1}{2}(1+ \alpha_{r_{w}})$ over one lattice
site $r_{w}$ in every row $j$. The symbol $r^{*}_{jw}$ denotes the 
first leftmost white lattice site $r_{w}$ in the $j$th row.
In combination with Eq.(\ref{Ihalpha}) [valid for any plaquette],
the conditions $\alpha_{r^{*}_{jw}} =1$ and $I_{h} =1$ ensure
that $\alpha_{r} =1$ for all $r$. 
A derivation of Eqs.(\ref{fermion_vac}, \ref{brw}) is given in the 
appendix. It is noteworthy that 
\begin{eqnarray}
~[I_h, \mathcal{B}_{r_{w}} ] = [I_h,I_{h'}] 
= [\mathcal{B}_{r_{w}}, \mathcal{B}_{r_{w}'}]
= [I_h, d_{k}^{\dagger}] \nonumber
\\ = [I_{h}, \overline{\alpha}_{r^{*}_{jw}}]= 
[\mathcal{B}_{r_{w}}, \overline{\alpha}_{r^{*}_{jw}}]= 0.
\end{eqnarray} 
$\{I_h\}$ and $\{\mathcal{B}_{r_{w}}\}$ 
lead to disjoint
$Z_{2}$ algebras. Thus, when combined, 
Eqs.(\ref{eD}, \ref{ks}, \ref{soln2kit2}, 
\ref{uv}, \ref{dkf}, \ref{fermion_vac},
\ref{brw}, \ref{alphaspin}) 
give us the explicit form of the ground state wavefunctions 
for the $S=1/2$ system defined by Eq.(\ref{H}).

\subsection{Comparison to ground states of other topological ordered
systems}

The spin basis form of the ground state of this topologically
ordered system can be compared to that of other
systems exhibiting topological quantum order.
E.g. both the Rokhsar-Kivelson [RK] \cite{RK} and the 
Kitaev toric code models \cite{Kitaev2003} (as well
as Wen's plaquette model \cite{wen_plaq} which is equivalent to 
Kitaev's toric code model) \cite{NO} have ground
states of the form
\begin{eqnarray}
|\psi \rangle = {\cal{N}} \sum_{i} | \chi_{i} \rangle
\label{topstate}
\end{eqnarray}
with ${\cal{N}}$ a normalization constant and the sum
over $i$ performed over all states $|\chi_{i} \rangle$
which belong to a given topological sector which 
is evident by the real space representation
(e.g. an even or odd number of dimers/positive bonds
in the Rokhsar-Kivelson/Kitaev toric code models
respectively). In both the RK and Kitaev toric code 
models, these 
ground states may be expressed 
as a product of projection operators 
(related to a $Z_{2}$ algebra)
acting on a given reference state.
For instance, in the Kitaev's toric code 
model on the square lattice \cite{Kitaev2003}
\begin{eqnarray}
H = - \sum_{s} A_{s} - \sum_{p} B_{p} 
\end{eqnarray}
with $A_{s} = \prod_{i} \sigma^{x}_{is}$
and $B_{p} = \prod_{\langle ij \rangle \in p} \sigma^{z}_{ij}$,
as stated earlier,  all ground states may be expressed as
\begin{eqnarray}
|g_{\mbox{toric~code}} \rangle &=&  
{\cal{N}} \left(\prod_{s} \frac{1}{2} (1+ A_{s}) \right)  
\nonumber\\ &&\times
\left( \prod_{p}  \frac{1}{2}
(1+ B_{p}) \right) | \phi \rangle,
\label{toriccodes}
\end{eqnarray}
with $| \phi \rangle$ a reference state.
Similarly, the general Eq.(\ref{topstate}) can 
be written as a projection of a given reference
state over a sector of Hilbert space that has a
chosen set of topological numbers (those corresponding
to a given topological sector).
As seen from our solution of Eq.(\ref{soln2kit2}), the ground
state of Kitaev's model on the hexagonal lattice (Eq.\ref{H}) \cite{Kitaev2006}
is more complicated. This is so as 
it involves the fermionic operators of Eq.(\ref{dkf}).
The Fermi vacuum of Eq.(\ref{fermion_vac}) indeed already 
has a form similar to Eq.(\ref{toriccodes}) but, as seen from 
Eq.(\ref{soln2kit2}), it needs to be 
acted on by Fermionic string operators.
This leads to a far more nontrivial state.
Written longhand in the original spin basis, 
we have that the fermion pair creation operator
\begin{eqnarray}
 d_{k}^{\dagger} d_{-k}^{\dagger} 
= 
 \frac{1}{4} \sum_{r_{w1}} \sum_{r_{w2}}  e^{i k (r_{w2} - r_{w1})}
\nonumber\\
\times  \left( \sigma^{y}_{r_{w1}} (\prod_{r'<r_{w1}}
\sigma^{z}_{r'}) - i \sigma^{z}_{r_{b1}} (\prod_{r'<r_{b1}} \sigma^{z}_{r'})\right) \nonumber
\\ \times \left(
\sigma^{y}_{r_{w2}}  (\prod_{r"<r_{w2}} \sigma^{z}_{r"}) - i \sigma^{z}_{r_{b2}} (\prod_{r"<r_{b2}} \sigma^{z}_{r"}) \right).
\label{explicitboson}
\end{eqnarray}
needs to be augmented. Here, in the product signs we made explicit
that along the Jordan-Wigner contour that we employ
here $r_{b} = r_{w} + L$.
When Eq.(\ref{explicitboson}) is combined with Eqs.(\ref{soln2kit2}, 
\ref{fermion_vac}, 
\ref{brw}), we have a form for the real space 
spin states of the Kitaev model.

\subsection{Time reversal symmetry breaking}

Our expressions show that the state $ |g \rangle$ 
of the open boundary condition system is generally 
not time reversal invariant for any size system-
whether it contains an odd or even number of spins. 
In the expressions above for the ground state
this is seen by taking $\sigma^{a} \to - \sigma^{a}$
for all components $a=x,y,z$ under time reversal. 
If any of the indices in Eq.(\ref{explicitboson}) 
are such that $(r_{w1}-r_{w2})$ or $L$
are odd then Eq.(\ref{explicitboson}) is not time
reversal invariant. Similarly, the operator $\mathcal{B}_{r_{w}}$
of Eq.(\ref{brw}) contains $(L+1)$ spin operators and is not
time reversal invariant for even $L$. A reference state
$| \phi \rangle$ of Eq.(\ref{fermion_vac}) 
on an even size lattice can be chosen
to be time reversal invariant. For the odd size lattice,
the considerations are more immediate.

\begin{figure}
\includegraphics[width=2.0in]{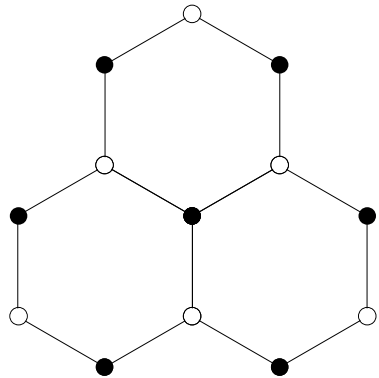}
\caption{The geometry of three hexagons that span a total 
of thirteen vertices. By Kramers' theorem,
ground states of this system that contains
an odd number of $S=1/2$ spins must break
time reversal invariance.}
\label{FIG-thirteen-vertices}
\end{figure}

As a concrete example, let us consider three hexagons that span a total of thirteen vertices as shown in Fig.\ref{FIG-thirteen-vertices}. 
This example serves to 
illustrate the presence of time reversal symmetry breaking 
even without doing any calculations.It is impossible to write down a ground state of Kitaev's model of thirteen S=1/2 spins of this 
thirteen site ``disk like geometry" that is non-degenerate. The same 
considerations apply also to instances in which we have various topologies. 
For instance, we can put the figure above of three hexagons with a total of 
thirteen sites on a three-fold torus-- that of three handles (the genus number
$g=3$). Here, there will be a hole at the center of each of the hexagons. 
One of the sites is common to three ordinary tori and three 
other sites are common to two ordinary tori. There is periodicity 
around each of the hexagons 
(hence the designation of $g=3$) in this case.  
The thirteen site system of Fig.(\ref{FIG-thirteen-vertices})
on the plane corresponds to a system with open 
boundary conditions while if it is a $g=3$ torus is corresponds to a system 
that has non-trivial periodicity around 3 hexagonal rings (along with three
other trivial periodicity directions of period one).
In both cases, as the Hamiltonian $H$ is time reversal invariant 
(because $T^{-1}S_{a;i}T = -S_{a;i}$ with $T$ the time reversal 
operator and $S_{a;i}$ the $a=x,y,z$ component of the spin operator 
at spin $i$ and the Hamiltonian is quadratic in the spins) and there 
is an odd number of spins ($n=13)$, 
by Kramers' theorem, the ground state-- and in fact any energy eigenstate-- 
must be, at least, two fold degenerate. General considerations for 
constructing a torus of genus $g$ from individual 
ordinary ($g=1$) tori are outlined in \cite{fusion}.
We emphasize that this degeneracy is mandated for an
odd size system by Kramers' theorem for any time reversal invariant
Hamiltonian. Time reversal symmetry breaking similarly
occurs in extensions of Kitaev's model
that allow for odd cycles \cite{Kitaev2006} \cite{steve} (e.g. a triangular
decoration of the lattice). 

In any half integer spin system 
in which the number of particles is odd,
the ground states must, by Kramers' theorem, 
exhibit time reversal symmetry
breaking. This conclusion may be fortified for general gapped systems
with short range Hamiltonians. In such systems, there is an exponential 
clustering of correlations. \cite{kohn} This clustering, in turn,
implies that the
ground state can, up to exponentially small corrections, \cite{hastingsfactor}
be written as a matrix product state constructed
of low lying states on finite size 
blocks. The block size may, in principle,
be taken arbitrarily large so long as it is 
smaller than the system. Consider a general Hamiltonan
$H = \sum_{R} h_{R}$
with $R$ labelling different blocks and $h_{R}$ a Hamiltonian that has its
support on $R$. Spins can be shared by different blocks $R$. By Hastings'
theorem \cite{hastingsfactor}, 
the ground state of $H$ is up to exponential corrections a matrix
product state. Let us now
examine what this implies when combined with time reversal. If all blocks $R$ 
are chosen to have an odd number of sites in the decomposition of any 
short range $S=1/2$ Hamiltonian $H$,
then all states of the different blocks $R$ will be, at least, doubly
degenerate. This is because of the (at least) two fold degeneracy of each
level in the odd size blocks $R$ implied by Kramers' theorem.
We see here that the matrix product 
construction then implies
that the ground state of large systems (also if they are of even
size) must also be, at least, two-fold degenerate up to 
corrections which are exponentially small in the size
of the system. This degeneracy is dictated by the time reversal non invariance
of the local energy eigenstates in each of the odd sized blocks $R$.
The considerations above are general and apply to both systems 
with or without topological 
order. Topological order pertains to
systems such as Kitaev's which donot display global symmetry
breaking associated with local order parameters. \cite{wenbook} 
The existence of degeneracy in Kitaev's model 
can also be seen by the non-commutativity of
existant symmetries: this non-commutativity
lies at the heart at Kitaev's inception
of this model.
We will detail symmetry
considerations for this model 
in a later section.

\subsection{Boundary terms}
For infinite open systems, 
the previous discussion will be sufficient for 
the analysis of bulk properties. However, in a closed system, 
boundary terms will lead to a topological dependency of the ground states. 
In this subsection, we shall consider the toridal geometry, which can be viewed as periodic boundary conditions along both directions. For the Jordan-Wiger transformation defined in Eq.(\ref{JWc}), the periodic boundary condition along the vertical direction will has no effect since the phase terms of two nearest neighboring terms along $x$ or $y$ bonds cancel out and there is no phase term for the $z$ bond coupling. For the periodic boundary condition, the boundary terms read
\begin{eqnarray}
H_{boundary}=\sum_{j}\left[J_x \sigma_{1,2j}^x\sigma_{L_h,2j}^x
+J_y\sigma_{1,2j+1}^y\sigma^y_{L_h,2j+1}\right].
\end{eqnarray}
Here, $L_h$ is the size of the sytem along the horizontal direction. After the Jordan-Wigner transformation (\ref{JWc}), the coupling strength of the boundary terms aquires a phase 
\begin{eqnarray}
H_{boundary}&=&\sum_{j}\left[J_x \phi(2j) A_{1,2j}A_{L_x,2j}\right.\nonumber\\
&\quad&\left. +J_y \phi(2j+1)A_{1,2j+1}A_{L_x,2j+1}\right]
\end{eqnarray}
with the phase term given by
\begin{eqnarray}
\phi(j)&=&\prod_{1\leq i\leq L_{h}}\sigma^z_{i,j}.
\label{pts}
\end{eqnarray}
This phase does not commute with the $Z_2$ bond operator $\alpha$ on the $z$-bonds attached to $j$-th horizontal line. The phase factor $\phi(j)$ reverses all $\alpha$ of the $z$-bonds attached to the sites on the 
$j$-the horizontal line.  The boundary terms thus lift the degeneracy characterized by different choices of $\alpha$ bonds. The gauge freedom discussed in section \ref{section-symmetry} is now fixed by the boundary terms and the ground state degeneracy acquires a topological dependence. On the other hand, the boundary term commutes with the plaquette quantity defined by Kitaev. Therefore, the discussion about the correlator in later section is still valid when boundary term is included. The momentum
components $(k_{x}, k_{y})$ employed in the solution of the previous 
subsections above are related by a 45 degree rotation to the discrete
values $(k_{h}, k_{z})$ for the system with periodic boundary
conditions along the vertical and horizontal directions
discussed here.

We conclude with brief remarks about the symmetries
of Eq.(\ref{pz}) in the case of periodic boundary conditions
and on a candidate trial state. The spin system 
of Eq.(\ref{H}) on the torus has the $\phi(j)$ 
operators of Eq.(\ref{pz}, \ref{pts})
as exact symmetries.
On the torus, the symmetry of Eq.(\ref{pz}) 
corresponds to the product
of spin operators along an entire toric cycle.
These operators satisfy the following identity
\begin{eqnarray}
\phi(j) \phi(j+1) = \prod_{h \in S} I_{h}.
\label{ptribbon}
\end{eqnarray}
In Eq.(\ref{ptribbon}), $S$ is a ribbon of width one
on the torus which contains all hexagons (plaquettes) lying between
the two consecutive rows $j$ and $j+1$. 
As the ground states of the Kitaev model are vortex free ($I_{h}=1$
for all $h$), the righthand side of Eq.(\ref{ptribbon})
is one. Thus, for all rows $j$, we have that
$\phi(j) = \phi(j+1)$. A variational state for
the periodic system is given by
\begin{eqnarray}
|\psi_{var} \rangle = {\cal{N}} \sum_{\alpha'} |g_{\alpha'} \rangle 
= {\cal{N}} \Big( \prod_{j} [1+ \phi(j)] \Big) |g \rangle
\label{varstate}
\end{eqnarray}
with ${\cal{N}}$ a normalization constant, 
and $\{\alpha'\}$ all 
configurations, which up to the symmetry operations of
 Eqs.(\ref{uiw}, \ref{pz}), correspond to the uniform
sector $\alpha_{r} =1$ for all bonds $r$. Here, $|g_{\alpha'} \rangle$
is the ground state in a given $\alpha'$ sector of the open
boundary system. Such a state
is related to the general uniform $\alpha_{r}=1$ ground state
which we derived earlier (the state 
$|g \rangle$ of Eqs.(\ref{soln2kit2}, \ref{uv},
\ref{dkf}, \ref{fermion_vac}, \ref{brw}))
by the application of the symmetry 
operators $\phi(j)$ of Eqs.(\ref{pz}, \ref{pts}). 
The variational state of Eq.(\ref{varstate}) is
an eigenstate of all of the symmetries $\{\phi(j)\}$ of Eq.(\ref{pz}, 
\ref{pts}) [with a uniform eigenvalue
which is equal to one].

\section{Anyons in the gapped phase}\label{section-anyon}
We will now derive and study anyons in the gapped phase
of Kitaev's model by relying directly on the Jordan-Wigner 
transformation. The existence of anyons in this phase has long
been recognized \cite{Kitaev2006} and, due to the prospect
of fault tolerant quantum computing, is
one of the main motivations for studying this system. 
The current appendix illustrates how anyons these may be 
directly studied and derived within our framework.

\begin{figure}
\includegraphics[width=2.8in]{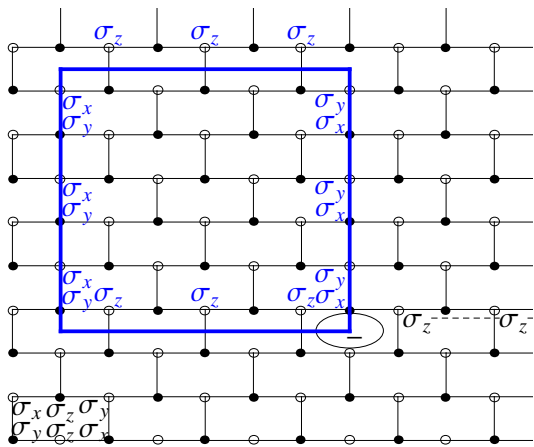}
\caption{Loop operator to move a vortex along a loop.}\label{FIG-anyon}
\end{figure}

Besides the Bogliubov quasiparticles, other excitations- the vortex 
excitation illustrated in Fig.\ref{FIG-states}(b)- are also manifest.
 It is more interesting to study these excitations within the gapped state. In this section, we shall follow Kitaev's original argument\cite{Kitaev2006} 
and now demonstrate, in our fermionic representation,
the anyonic nature of these vortex excitations. Let us start in the limit where $J_x=J_y=0$. In this limit, the fermionized Hamiltonian reads
\begin{eqnarray}
H=J_z\sum_r \alpha_r (2d^\dag_r d^{}_r-1).
\end{eqnarray}
The ground state is thus $2^{N/2}$ degenerate, where $N$ is number of sites. This degeneracy can also be understood in the original spin language. In this limit, all vertical bonds are disconnected. Let us consider a single bond, $(J_z\sigma^z_1 \sigma^z_2)$. Without loss of generality, let us consider $J_z>0$. There are two degenerate ground states $|\uparrow\downarrow\rangle$ and $|\downarrow\uparrow\rangle$. 
This degeneracy is related to the local $Z_2$ symmetry which present for $J_xJ_y=0$. Without losing generality, let us set $J_x=0$. In this case, we can define a local unitary transformation which is just $\sigma^x_1 \sigma^x_2$ on one of the $x$-bonds. The effect of this operator is to transform $J_z$ to $-J_z$ on the two vertical bonds connected to the $x$-bond. If $J_x J_y\neq 0$, such transformation is impossible and we expect the degeneracy to be lifted. The lowest order contribution from $J_x$ and $J_y$ term is thus expected to be in the second order of $J_xJ_y$, {\it i.e.}, $J_x^2J_y^2/J_z^3$. This corresponds to a fourth-order perturbation. Indeed, Kitaev has shown this is true in the language of spin operators\cite{Kitaev2006}. The non-constant effective Hamiltonian up to  $4$-th order is thus proportional to the product of the four horizontal bonds of a given plaquette, $(\sigma^x_1\sigma^x_2)(\sigma^y_2\sigma^y_3)(\sigma^y_6\sigma^y_5)(\sigma^x_5\sigma^x_4)$. In the language of our Majorana fermions $A$, this is 
\begin{eqnarray}
H_{eff}^{(4)}&\propto& (A_1A_2)(A_2A_3)(A_6A_5)(A_5A_4)\nonumber\\
&=&(iA_1A_6)(iA_3A_4).
\end{eqnarray}
Here, we have dropped the sublattice subscripts $b$ and $w$ for the $A$ fields. If we introduce a dual spin on vertical bonds, 
\begin{eqnarray}
\mu^z_{r}=2d^\dag_r d^{}_r-1,
\end{eqnarray}
then the effective Hamiltonian $H_{eff}$ will be nothing but an Ising coupling between two neighboring spins along horizontal direction $H_{eff}^{(4)}\propto \mu^z_{i,j+1/2}\mu^z_{i+2,j+1/2}$. 
The total perturbative Hamiltonian thus reads
\begin{eqnarray}
H=\sum_{ij}\left[2J_z \alpha_i \mu^z_i - 
\frac{J_x^2J_y^2}{16J_z^3}\mu^z_{ij}\mu^z_{i+2,j}\right]+const.
\end{eqnarray}
on the lattice of vertical bonds. The coefficient $-1/16$ is worked out by Kitaev\cite{Kitaev2006}. As one would expect, the $J_x$ and $J_y$ terms lift the degeneracy and the ground state is the one with $\alpha_i=1$ in our aforementioned gauge fixing of the extra freedom in section \ref{section-symmetry}. The interesting excitation state that corresponds to the anyon studied by Kitaev\cite{Kitaev2006} is thus the ground state of the Hamiltonian of the configuration sketched in Fig.\ref{FIG-states}(b). This is exactly the same vortex state in plaquette $p$ defined by Kitaev\cite{Kitaev2006} where one conserved quantity $I$ is reversed in the plaquette $p$. In this state, to save (positive)
$J_z$ contributions to the energy, the effective spins $\{\mu_i\}$ on the cut where $\alpha_i=-1$ are also flipped and 
the resultant configuration is a domain wall structure at $p$. The energy penalty associated with this
domain wall is$J_x^2J_y^2/8J_z^3$. 
For convenience let us define

\begin{eqnarray}
\tilde{I}(i,j+1/2)=\mu^z_{i-1,j+1/2}\mu^z_{i+1,j+1/2}.
\end{eqnarray}
In terms original spin operators, $\tilde{I}$ is, similar to Eq.(\ref{Eq-kitaev-conserved}), 
\begin{eqnarray}
\tilde{I}=\sigma^x_1\sigma^z_2\sigma^y_3\sigma^x_4\sigma^z_5\sigma^y_6,
\end{eqnarray}
An illustration is provided in the lower corner plaquette of Fig.\ref{FIG-anyon}.

We shall now demonstrate that the anyonic nature of the vortex excitation.
A vortex on the plaquette centered at $(i,j+1/2)$ is characterized by two kinks on the same bond, one in the channel of $\alpha$ and another in the channel of  $\mu^z$. There are four equivalent ways to create a vortex on the plaquette centered at $(i,j+1/2)$:
\begin{eqnarray}
P_w^{R}(i,j+1/2)&=&\prod_{i'>i} iA_w B_w(i',j+1)\\
P_w^{L}(i,j+1/2)&=&\prod_{i'<i} iA_w B_w(i',j+1)\\
P_b^{R}(i,j+1/2)&=&\prod_{i'>i} iA_b B_b(i',j)\\
P_b^{L}(i,j+1/2)&=&\prod_{i'<i} iA_b B_b(i',j).
\end{eqnarray}
The operators $P_w^R$ ($P_b^R$) are related to $P_w^L$ ($P_b^L$) by the gauge transformations encapsulated by 
$U^w_i$ ($U^b_i$) of Eq.(\ref{uiw}). 
We also notice that $\left(P_{b,w}^{R,L}\right)^2=1$. Therefore, $P_{b,w}^{R,L}$ can either create a vortex or annihilate an existing vortex at $(i,j+1/2)$. A naive way to move an anyon vertically from $(i,j+1/2)$ to $(i,j+5/2)$ is 
to apply $P_b^{L}(i,j+5/2) P_w^{L}(i,j+1/2)$. It turns out that it is more convenient to multiply two extra phase terms, the first one contains the multiplication of $z-bonds$ centered at $(i',j+3/2)$ with $i'<i$ 
\[\left(\prod_{i'<i}[(iA_bB_b)(iA_wB_w)]_{z-bonds}\right),\]
and the second one contains the multiplication of $\tilde{I}$ centered at $(i',j+3/2)$ with $i'\leq i$, $\prod_{i'\leq i}\tilde{I}(i',j+3/2)$. After including these two phases, the shifting operator $\mathcal{T}_y$ from $(i,j+1/2)$ to $(i,j+5/2)$ takes a simple form in terms of spin operators
\begin{eqnarray}
\mathcal{T}_y = \sigma^x(i,j+1)\sigma^y(i,j+2).
\end{eqnarray}

We are now ready to construct the loop operator that moves vortex around and  study the statistic properties of vortex excitations. Let us first create a vortex on the plaquette centered at $(i,j+1/2)$ with $P_w^R(i,j+1/2)$, which applies $S^z$ on all site-$3$ (white site at top-right corner) of the plaquettes $(i',j+1/2)$ with $i'\geq i$ as shown in Fig.\ref{FIG-anyon}. To move this vortex along a closed loop $c$, we can construct a a loop operator $L_{C}$ by multiplying $\mathcal{T}_y$ along vertical lines and $\sigma^z$ along horizontal parts of the loop, as illustrated in Fig.\ref{FIG-anyon}.
The loop operator can be expressed by 
\begin{eqnarray}
L_{C}=\left(\prod_{\substack{(i',j')\in C,\\ j'+j=odd}}\tilde{I}(i',j'+1/2)\right)
\left(\prod_{l_b\in C}\left[\sigma^z_{w}\sigma^z_{b}\right]_{l}\right).
\end{eqnarray} 
An explanation of the notation is due. Without losing generality, let us assume the anyon is located on an even row, $j=2n$. $L_C$ can be separated into two parts. The first part is the product of $\tilde{I}$ that covers the plaquettes on the odd rows enclosed by the loop $C$. The second part is the product of $\sigma^z_b\sigma^z_w$ on the vertical bonds whose black site is enclosed inside the loop $C$. In the projected subspace on which the effective Hamiltonian lives, the vertical bond part is the identity $1$. Therefore, the loop operator is just the product of $\tilde{I}$ over the plaquettes on the odd rows enclosed by $C$, $L_{C}=(-1)^{n_C}$ where $n_C$ is the number of vortices that are located on the odd rows and inside the loop $C$. In another words, if we bring a vortex on odd row around another vortex on an even row, an extra minus sign is generated, while no such minus sign if we bring a vortex on an even (odd) row around another vortex also on an even (odd) row. 
Putting all of the pieces together, we fleshed out, by building on the fermionic
representation, the anyonic nature of the vortex excitations.

\section{String and brane type 
correlators dictated by local symmetry}\label{section-correlators}

In this section, we shall apply symmetry considerations
\cite{NO} \cite{BN} and review condensed remarks from \cite{BN} concerning
the symmetries of this model and general considerations regarding
non-vanishing correlators in similar matter coupled theories
which are reviewed in the Appendix (\ref{app2}).
We will show directly from symmetry considerations that the Kitaev
model on the hexagonal lattice \cite{Kitaev2006} does not exhibit any
two point correlations of length larger than one and that
any non-vanishing correlation function generally amounts
to a string operator (either closed or open). The fermionization
that we employ here \cite{Chen2007} \cite{Feng2006} 
gives rise to precisely such string like correlators
whose form is dictated by symmetry. In the
matter coupled gauge analogy of Section(\ref{app2}), all
possible correlators correspond to
open meson or closed photon lines. 
We will show that Kitaev's model supports finite 
valued {\em brane} type correlators.
By {\em brane} correlators, we allude to
correlators which span portions of the 
two dimensional plane. This result- as it 
applies for only the two point correlations- was also recently derived by 
a Majorana fermion construction \cite{bms}. In
what follows, we show how gauge symmetry considerations
effortlessly mandate this result. 

\subsection{Two point correlators}

We claim
that both at finite and at zero 
temperature the spin-spin correlation between any
two spins which are  separated by more than 
one lattice constant must vanish. To
see this,  we note that this system 
displays the local symmetries of Eq.(\ref{Eq-kitaev-conserved}).
We now fuse these symmetries together and see that
we have a symmetry associated with any closed contour $C$, by 
\begin{eqnarray}
\hat{O}_{C} = \prod_{p \in C} \sigma_{p}^{\gamma_{p^{C}}}.
\label{oc}
\end{eqnarray}
In Eq. (\ref{oc}), the polarization index $\gamma_{p^{C}}$ is chosen to
correspond to the direction ($\gamma = x,y,z$) of the single bond
emanating from site $p$ which  does not lie in the contour $C$. Any
given site $p$  forms the endpoint of three bonds; two of these bonds
lie in the contour $C$ and only a single  bond ending in $p$ does not
lie in $C$. 

We now briefly relate these symmetries to degeneracies.
If two curves $C$ and $C'$ share an odd
number of sites in common then $O_{C}$ and $O_{C'}$ anticommute.
Thus if, e.g, we choose to simultaneously diagonalize $H$ and $O_{C}$
and determine a ground state $|g_{1} \rangle$ in that common eigenbasis, 
then $O_{C'}|g_{1} \rangle$ is a new degenerate ground state. 
This degeneracy applies not only to the ground
state sector but to all energy levels. 

\subsubsection{$T>0$} 

By employing these symmetries along with  Elitzur's 
theorem \cite{Elitzur} at temperatures $T>0$, we find that the 
finite temperature  correlator $\langle
\sigma_{p}^{a} \sigma_{q}^{b} \rangle$ vanishes unless (1) $p$ and $q$
are nearest neighbor sites and that (2) $a=b$ corresponds to the
direction ($a=x,y,z$) between the two lattice sites $i$ and $j$ in
question. This is so as unless conditions (1) and (2) are satisfied,
there is at least one loop $C$ for which  $[\sigma_{p}^{a}
\sigma_{q}^{b}]$ is not invariant under $\hat{O}_{C}$ of Eq. (\ref{oc})
and for which 
\begin{eqnarray}
\hat{O}_{C}  [\sigma_{p}^{a} \sigma_{q}^{b}] \hat{O}_{C}
= - [\sigma_{p}^{a} \sigma_{q}^{b}].
\end{eqnarray}

\subsubsection{T=0} 
The $T=0$ relation (even for different times \cite{bms})  is similarly
derived. All of the operators $\hat{O}_{c}$ commute with one another.
[This is seen as for the minimal loop $C$ which contains a single
hexagon, two neighboring hexagons leading to operators $\hat{O}_{1}$ and
$\hat{O}_{2}$ which contain two common sites which lead to two 
anti-commutation relations between the different Pauli matrices.
\cite{Kitaev2006}. The product of operators each acting of a single 
hexagon leads to the most general symmetry operator of Eq. (\ref{oc})
with an arbitrary path $C$.] Consequently, we may simultaneously
diagonalize $H$ of Eq. (\ref{H}) with all of the operators
$\hat{O}_{c}$. The dynamical $T=0$ two spin correlation function
\begin{eqnarray}
&&\langle \psi | \sigma_{p}^{a} (0) \sigma_{q}^{b}(t) | \psi \rangle \nonumber
\\ &=& 
\langle \psi | \sigma_{p}^{a} (0) \exp(i H t) \sigma_{q}^{b}(0)  
\exp(-i Ht) | \psi \rangle \nonumber
\\ &=& \Big( \langle \psi | \sigma_{p}^{a} \sigma_{q}^{b} | \psi \rangle
+ i t \langle \psi | \sigma_{p}^{a} [H, \sigma_{q}^{b} ] | \psi 
\rangle \nonumber
\\ &&+ \frac{(it)^{2}}{2!} \langle \psi | \sigma_{p}^{a}
[H,[H,\sigma_{j}^{b}]] | \psi \rangle + ... \Big)
\label{trivialexp}
\end{eqnarray}
with $| \psi \rangle$ any ground state which  is a simultaneous
eigenstate of $H$ and $\hat{O}_{C}$. The eigenvalues of $\hat{O}_{C}$
can only be $\pm 1$. We can now show that unless $p$ and $q$ are
linked by a single step along a direction $a$ each of the
expectation values in Eq. (\ref{trivialexp}) vanishes. For example, for the
first term in the final expression, we have
\begin{eqnarray}
\langle \psi | \sigma_{p}^{a} \sigma_{q}^{b} | \psi \rangle 
=(\langle \psi | \hat{O}_{C})  
\sigma_{p}^{a} \sigma_{q}^{b}  (\hat{O}_{C} | \psi \rangle)  \nonumber
\\= \langle \psi |( \hat{O}_{C} \sigma_{p}^{a} \sigma_{q}^{b} \hat{O}_{C}) 
| \psi \rangle  = 
- \langle \psi | \sigma_{p}^{a} \sigma_{q}^{b} | \psi \rangle,
\label{st0}
\end{eqnarray}
unless  $p$ and $q$ which are linked by a single step along a direction
$a$. Similarly, the second term in Eq. (\ref{trivialexp}) is seen to
vanish. Here, the commutator $[H, \sigma_{q}^{b} ]$ leads to sums of
bilinears which involve the lattice site $j$ and a nearest neighbor
site $k$. When multiplied by $\sigma_{p}^{a}$ this leads an expression
which is cubic in the spin operators. If all of the three lattice
sites  are different, we can choose a path $C$ such that this
expression changes sign under a unitary transformation corresponding to
$\hat{O}_{C}$. Similar higher order terms in Eq. (\ref{trivialexp})
unless $p$ and $q$ which are linked by a single step along a direction
$a$. This result reaffirms (from a purely symmetry point of view) the
more detailed derivation of \cite{bms}. 

\subsection{Higher order string and brane type correlators}
\label{hior}
We now turn to higher order spin correlations.
A moment's reflection reveals that
invariance under the general
local symmetries of Eq.(\ref{Eq-kitaev-conserved})
allows only for the string or {\em brane} operators
(either closed or open) to be finite: 
in other words, all spins must form continuous 
clusters along lines (strings) or lie 
in a fragment of the
two dimensional plane.

String and {\em brane} type 
correlators may involve, in the thermodynamic limit,
an infinite number of fields and allow different topologies
than that of closed loops alone. The fields may be ordered
along an open string (or collection of
such strings) or in higher dimensions may involve
sophisticated combinations of fields at all lattice sites
which we term as {\em brane} type correlators. 

The best known example of such a string correlator
is that in the spin $S=1$ AKLT chain \cite{AKLT88} \cite{KT}  
in which 
\begin{eqnarray}
\langle S_{i}^{z} \Big( \prod_{j=i+1}^{k-1}  e^{i \pi S_{j}} \Big) S_{k}^{z} \rangle = \frac{4}{9}.
\label{AKLT}
\end{eqnarray}

Eq.(\ref{AKLT}) holds for arbitrarily far separated sites $i$ and $k$.
As this correlator is, asymptotically, far larger than the usual
spin-spin correlator, it is often said to capture a hidden
order. We may construct string operators
which are more sophisticated variants of the operator
appearing in Eq.(\ref{AKLT}). Apart from spin chains, \cite{AKLT88}
\cite{KT} \cite{Nijs}
such correlators also appear,
amongst others, in doped Hubbard chains and related systems, \cite{Jurij}
\cite{OS} \cite{KMNZ} cold atom chains \cite{coldatomstring}, 
in spin leg ladders \cite{todo}, and may be related 
to non-local constructs in Quantum
Hall systems \cite{GM}. In doped Hubbard chains, 
the string correlator decays asymptotically with
distance but with a power which is smaller
than that of the usual spin-spin correlators.
To date, nearly all appearances of 
string correlators are confined to 
one dimensional or quasi-one dimensional
systems. The only known exceptions 
which we are aware of are \cite{Jurij} and 
possible links to the density matrix construct of
\cite{GM} for Quantum Hall systems. 
In what follows, we show 
that the Kitaev model is a rigorous 
two dimensional example of a system
in which various string correlators can 
either attain a finite value or
decay asymptotically in a slow 
manner (explicitly so for gapless systems).

Let us turn back to Kitaev's model of 
Eq.(\ref{H}). 
The only gauge invariant (symmetry of Eq.(\ref{Eq-kitaev-conserved}))
quantities which, by Elitzur's theorem \cite{Elitzur} can
attain a finite expectation value are given by 
\begin{eqnarray}
\left\langle  \prod_{pq \in C} \left[\sigma_{p}^{\alpha_{pq}} 
\sigma_{q}^{\alpha_{pq}}\right] 
\right\rangle
\label{ocopen}
\end{eqnarray}
with $C$ being a general contour 
which may be open or closed 
(or a union of such contours)
and $\alpha_{pq}$ denotes the direction of the physical bond between 
two nearest neighbor sites $p$ and $q$. When $C$ is a closed
loop, the argument of the average in Eq.(\ref{ocopen}) is, 
up to a multiplicative phase factor, equal to
the symmetry operator of Eq.(\ref{oc}).

\begin{figure}
\includegraphics[width=2.8in]{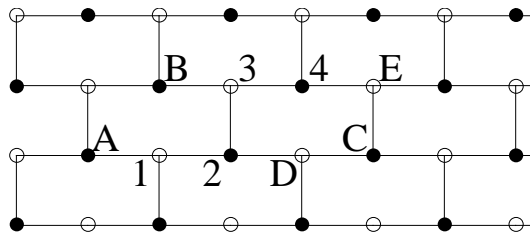}
\caption{An open loop correlator of the form
given by Eq.(\ref{ocopen}). Here, the correlator is
equal $ \langle 
(\sigma^{x}_{1} \sigma^{x}_{2})(\sigma^{z}_{2} \sigma^{z}_{3}) 
(\sigma^{x}_{3} \sigma^{x}_{4}) \rangle
= \langle \sigma^{x}_{1} \sigma^{y}_{2} 
\sigma^{y}_{3} \sigma^{x}_{4} \rangle$. All correlators
are of the form of continuous string correlators such
as this or of their union. See Eq.(\ref{ocopen}). As
we review in the appendix, precisely
such string correlators are the only non-vanishing 
correlation functions in matter coupled gauge theories.}
\label{FIG-loop}
\end{figure}

An example of such a correlator is furnished in 
Fig.(\ref{FIG-loop}). In Fig.(\ref{FIG-loop}), we show
the only non-vanishing correlator which has
four spins with sites 1 and 4 as its endpoints.
To see that this string correlator
is invariant under all local symmetries, consider first sites 1 and 4.
Considering hexagons $h=E,D$ (see figure), we see that
only the fields $\sigma^{x}_{1}$ and $\sigma^{x}_{4}$
are invariant under 
the local symmetry of Eq.(\ref{Eq-kitaev-conserved}).
Next, we note that this particular choice
of these fields at sites 1 and 4, 
the symmetry of Eq.(\ref{Eq-kitaev-conserved})
for both hexagons $h=A$ and $h=B$, enables
the introduction of a field at sites 2,3 
of the form $\sigma^{y}_{2,3}$. Thus, 
$\langle \sigma^{x}_{1} \sigma^{y}_{2} \sigma^{y}_{3} \sigma^{x}_{4} \rangle$
is a correlator which is invariant under all local symmetries.
As such, it is not prohibited from attaining a non-zero expectation value
at finite temperatures by Elitzur's theorem. 
Similarly, the product of two disjoint bonds
$\langle (\sigma_1^{x} \sigma_{2}^{x})( \sigma^{3}_{x} \sigma^{4}_{x}) 
\rangle$ is invariant under all local symmetries.
In a similar fashion, we can proceed to consider 
longer contours invariant under local symmetries-
all of which must be of the form of Eq.(\ref{ocopen}).
Replicating the $T=0$ considerations of Eq.(\ref{st0})
to multi-spin correlators, we see that
string correlators of the form of Eq.(\ref{ocopen})
are similarly symmetry allowed within the ground state sector
but others are not. By construction, the string correlator 
is invariant under all products
of the local symmetries $\{I_{h}\}$-  
the symmetries of Eq.(\ref{oc}).
The contour $C$ may consist of
disjoint open segments (e.g. disjoint bonds with each bond
containing two sites). Of all of the correlators of Eq.(\ref{ocopen})
that we found to be allowed by Elitzur's theorem, only those with
an even number of spins are time reversal
invariant. String correlators of the form of 
Eq.(\ref{ocopen}) with an odd number of
spins are not time reversal invariant and 
consequently must vanish when time 
reversal symmetry is unbroken. \cite{julien}

For closed contours $C$,
the quantity to be averaged in Eq.(\ref{ocopen}) becomes none other
than the symmetry of Eq.(\ref{oc}). In Appendix (\ref{app2}), 
we review similar selection rules for lattice gauge 
theories. Though all non-vanishing correlators
must be of the form of Eq.(\ref{ocopen}), systems
with open boundary conditions, further allow for additional 
string correlators of this form with only the boundary site(s)
on the contour $C$ not adhering to this form.
The appearance of only such non-vanishing open or
closed continuous string correlators (involving
all sites between sites $p$ and $q$) is reminiscent of
the gauge invariant correlators appearing
in matter coupled gauge theories, e.g. \cite{BN}. 
As stated earlier, all non-vanishing correlators
must be of the form of Eq.(\ref{ocopen}) on a system
with periodic boundary conditions or, in a system
with open boundary conditions, further allow for additional 
string correlators of this form with only the boundary site(s)
on the contour $C$ not adhering to this form.
The ideas underlying the constructs which 
we introduce next were reviewed in
\cite{NO}.

\subsubsection{Correlators of maximal value} 

We next construct several such string correlators
by relying on the mapping to the Fermi problem.
This mapping transforms correlators which
involve only several Fermi fields
into those which involve an extensive
number of spin fields and vice versa.

In the Kitaev model, the Fermi vacuua states $| 0 \rangle$ can be chosen 
to simultaneously
diagonalize all of the symmetry operators $\{I_h\}$
and $\{\mathcal{B}_{r_{w}}\}$. Here, we will have that
\begin{eqnarray}
\left|\langle 0| \prod_{r'=r_{w1}}^{r_{w2}} 
B_{r'}|  0 \rangle \right|
&=& \left|\langle 0
| \sigma^{x}_{r_{w1}} \sigma^{y}_{r_{w1}+1} \sigma^{x}_{r_{w1}+2} ...
\sigma^{x}_{r_{b2}-1} \sigma^{y}_{r_{b2}} |0 \rangle\right| \nonumber
\\ &=& 1
\label{string}
\end{eqnarray}
(for an even number of sites). 
The string correlator of Eq.(\ref{string}) is equal to a constant value
irrespective of its length (the number of spin operators that
it contains). Different Jordan-Wigner contours may also be chosen 
which lead to paths other than those in Eq.(\ref{string}) which
connect the two endpoints. The Fermi vacuua become ground 
states only in the limit of $|J_{z}/J_{x,y}| \gg 1$
(and permutations thereof). Physically in this limit, 
the large value of $|J_{z}|$ forces consecutive chains 
to be fully correlated. The open string correlator
of Eq.(\ref{string}) augments the
closed loop correlators of Eq.(\ref{oc})
for any closed contour $C$. 
Of course, combinations of various closed loops and
open string operators are possible.

We now discuss the situation for arbitrary $J_{x,y,z}$. 
Given a general ground state $| g \rangle $ of the form of
Eq.(\ref{soln2kit2}) we can transform it
into a Fermi vacuum state by a unitary transformation
\begin{eqnarray}
U = \prod_{k} \left(u_{k} + v_{k}(b_{k}^{\dagger} + b_{k}) - u_{k} n_{k}\right), 
\quad U | g \rangle = | 0 \rangle.
\end{eqnarray}
Here, $n_{k} \equiv d_{k}^{\dagger} d_{k}$. 
For a general ground state $|g \rangle$
the non-local {\em brane} expectation value
\begin{eqnarray}
|\langle g| U^{\dagger}  \sigma^{x}_{r_{w1}} \sigma^{y}_{r_{w1}+1} 
\sigma^{x}_{r_{w1}+2} ...
\sigma^{x}_{r_{b2}-1} \sigma^{y}_{r_{b2}}  U | g \rangle| =1.
\label{maximal}
\end{eqnarray}
The operator of Eq.(\ref{maximal}) has maximal correlations
(the modulus of the expectation value of these operators
cannot exceed 1). The maximal correlations
exceed the standard string correlators 
found in the AKLT chain. 
Written in terms of the original
spin fields at each site, this operator generally (for
finite $J_{x,y,z}$) spans all sites of the lattice.

\subsubsection{non-maximal correlators}

We can similarly construct other operators which would reduce
to other open string operators if the system were one dimensional
(if, e.g., $J_{z} =0$). These form other generalizations
of the familiar one dimensional string operators. In what follows,
we discuss string operators which are not maximally (and indeed
decay with increasing separation between the two endpoints of
the string. To this end, let us start by writing
\begin{eqnarray}
d_{r_{1}}^{\dagger} d_{r_{2}} 
&=& \pm \frac{1}{4}\left(\prod_{r_{w_{1}} \le r'''<r_{w_{2}}} \sigma_{r'''}^{z}\right) \nonumber
\\ &\times&
\left\{- i \sigma_{r_{b_{1}}}^{y}
+ \sigma_{r_{b_{1}}}^{x} \left(\prod_{r_{w_{1}} \le r' < r_{b_{1}}-1} \sigma^{z}_{r'}\right)
\right\}\nonumber
\\ &\times&  \left\{ i \sigma^{y}_{r_{w_{2}}}
+ \sigma^{x}_{r_{b_{2}}}\left(\prod_{r_{w_{2}} \le r''<r_{b_{2}}-1}  \sigma^{z}_{r''}\right)
 \right\}.
\label{dd}
\end{eqnarray} 
Here, the $+$ sign is chosen if $r_{w_{2}}<r_{b_{1}}$ along the
contour on which the Jordan-Wigner transformation is performed
and the $-$ sign is chosen otherwise.
Here, we have that
\begin{eqnarray}
\langle d_{r_{1}}^{\dagger} d_{r_{2}} \rangle 
= \frac{1}{2} \int \frac{d^{2}k}{(2 \pi)^{2}} 
e^{- i \vec{k} \cdot (\vec{r}_{1} - \vec{r}_{2})} 
\left[ 1 -  \frac{\epsilon_{k}}{E_{k}} \right],
\label{FT}
\end{eqnarray}
with the definitions of Eq.(\ref{ks}) for $\epsilon_{k}$ and $E_{k}$.
The form of this expression follows directly from 
the state of Eq.(\ref{soln2kit2}), $\langle n_{k} \rangle = v_{k}^{2}$.
Looking at Eq.(\ref{FT}), we see
that for gapless systems, a power law 
behavior may be sparked. In the presence
of a gap, the branch points in the complex
$q$ plane of $E_{q}$ 
of Eq.(\ref{ks})
along  a chosen direction may determine
the asymptotic long distance
correlation length along that
direction. Rather explicitly, in the gapped phase, we
have from Eq.(\ref{coeq}) that 
the logarithm of the correlator of Eq.(\ref{FT}) scales asympotically, 
along the $x,y$ directions
as $[-(|r_{2;a}-r_{1;a}|/\xi_{a})]$
(where $\{r_{a}\}$ are $\vec{r} \cdot \hat{e}_{a}$ with $a=x,y$, [see 
Fig.(\ref{FIG-hexagonal})]) 
with
\begin{eqnarray}
\frac{1}{\xi_{x}} = \cosh^{-1} \Big[
\frac{J_{x}^{2} + J_{z}^{2}-J_{y}^{2} }{2J_{x} J_{z}} \Big], \nonumber
\\ \frac{1}{\xi_{y}} = \cosh^{-1} \Big[ \frac{J_{z}^{2} 
+ J_{y}^{2}- J_{x}^{2}}{2J_{y} J_{z}} \Big].
\label{corli}
\end{eqnarray}

Along any given direction $a$, in the complex $q_{a}$ plane,
there are two branch points along the imaginary $q_{a}$ axis. 
There is a simple signature of the $T=0$ critical phase that occurs
when we cross from the gapped to the non-gapped phases
when $|J_{x} \pm J_{y}| = |J_{z}|$.
Here, the branch points merge at $q_{a}=0$ and 
there is a divergent correlation length $\xi_{a}$. Within the 
gapless phase, the two branch lie along 
the real $q_{a}$ axis.

Fusing Eqs.(\ref{dd}, \ref{FT}) together, we have
an expression for the expectation value of the string operator
which lives on all sites linking $r_{w_{1}}$ up to (and including)
$r_{b_{2}}$. The left hand side of Eq.(\ref{dd}) is given by the Fourier
transform in Eq.(\ref{FT}). 
Similar transformations may be applied to other
fermionic correlators which when translated into
the spin variables lead to other string correlators.

The ground state form of Eq.(\ref{soln2kit2})
enables a direct computation of the expectation
value of any general product of the form $\langle \prod_{i} d_{k_{i}}
\prod_{j} d^{\dagger}_{k_{j}} \rangle$. This is so as
the state of Eq.(\ref{soln2kit2}) is a direct product
in $k$ space. For each value of $k$, we have
a binomial distribution for each occupancy,
$v_{k}^{2}$ for the occupancy of the pair ($k, -k$)
and a probability of $u_{k}^{2}$ for it to be 
empty. Thus, any general multi-spin correlator
can be trivially computed- we first Fourier 
transform it, then express the product in
terms of the Fermi variables following a
Jordan-Wigner transformation, and then
compute the average for the states of Eq.(\ref{soln2kit2})
where each average for a given value of $k$ 
becomes decoupled from all other values of $k$.
In a related vein,
Wick's theorem for Eq.(\ref{ks})
ensures the decomposition of general
correlator into a product of pair
correlators for ($k,-k$) pairs.

\section{Conclusion}\label{section-conclusion}

In conclusion, we presented an exact solution
of Kitaev's toric code model which allows
for various new results:

({\bf 1}) We fermionize Kitaev's
model to render it into a 
p-wave type BCS pairing problem.

({\bf 2}) We derive the exact {\em ground state wavefunction
in terms of the spin degrees of freedom}. With it, we
examine the zero temperature correlation functions and derive
the exact correlation lengths within the gapped
phase. We 
further show that the wavefunction 
in this case is more
complicated than
in the other prototpyical
models of topological quantum 
order (e.g. the Rokhsar-Kivelson Quantum Dimer Model and
Kitaev's or Wen's square lattice models).

({\bf 3}) We prove that local symmetries only
enable string (open or 
closed) and {\em brane} type
correlators (and their unions) to be non-zero in this system.
We show how string  correlators may be directly evaluated
in terms of the exact ground state wavefunction. 
Kitaev's model is one of the very few two
dimensional systems with string or {\em brane} type correlators. 
Nearly all known examples of string 
correlators to date centered on one dimensional
systems.

({\bf 4}) We identify the local symmetries of
this system in terms of bond variables in the fermionic
problem. 

({\bf 5}) We illustrate, in terms of fermions, 
the anyonic character of the 
vortex excitations in the gapped phase 
by an explicit construction.

Many possible extensions of our results follow from our 
fermionic wavefunctions. For instance, we can consider
impurity bound states in systems engineered to
have Hamiltonians which deviate slightly
from Eq.(\ref{H}). For spatially non-uniform 
$J_{x},J_{y}, J_{z}$, so long as no new interactions are added, 
the toric symmetries remain in tact (as does
the fermionization). In the vortex free sector, the problem reduces
to that of an impurity in a BCS type system
which may lead to fermionic bound states. 
In subsequent work, we will further aim to detail 
the Ising type Hamiltonian that results for
the vortex variables by tracing out
over the fermionic degrees of 
freedom in Eq.(\ref{EQ-fermion-model}).

\section{Acknowledgments}

We thank J. Vidal for a careful reading 
of this work and remarks.

\appendix

\section{The Fermionic vacuum in the spin representation}
\label{app1}

In what follows, we briefly outline the derivation of
Eq.(\ref{fermion_vac}). For the fermionic vacuum 
we have $d_{k} |0 \rangle = 0$ for all $k$
or equivalently
\begin{eqnarray}
d_{r} | 0 \rangle =0
\label{drzero}
\end{eqnarray}
for all vertical bond centers $r$. Eq.(\ref{drzero}) implies that 
in the original spin basis, 
\begin{eqnarray}
\left[ \left( \prod_{r'< r_{b}} \sigma^{z}_{r'} \right) 
\sigma^{x}_{r_{b}} + i \sigma^{y}_{r_{w}} \left( \prod_{r'< r_{w}} 
\sigma^{z}_{r'} \right) \right] |0 \rangle =0.
\label{spinfermivac}
\end{eqnarray}
Now, let us write the state $| 0 \rangle$ 
as a general superposition of states in the $\sigma^{z}$
basis.
\begin{eqnarray}
|0 \rangle = \sum_{\sigma_{1} \sigma_{2} ... 
\sigma_{N}} A_{\sigma_{1} \sigma_{2} ... \sigma_{N}} | \sigma_{1} 
\sigma_{2} ... \sigma_{N} \rangle.
\label{0inSz}
\end{eqnarray}
Inserting Eq.(\ref{spinfermivac}) in Eq.(\ref{0inSz}) we find that
for all four possible spin orientations at $r_{w}$ and $r_{b}$
(up/down at each of these two sites) we have that the amplitudes
satisfy
\begin{eqnarray}
A_{\sigma_{1} \sigma_{2} ... \sigma_{r_{w}} ... \sigma_{r_{b}} ...
\sigma_{r_{N}}}  = 
-\left( \sigma_{r_{w}+1} ... \sigma_{r_{b}-1} \right) \nonumber
\\ A_{\sigma_{1} \sigma_{2} ... \sigma_{r_{w}-1} (-\sigma_{r_{w}}) 
\sigma_{r_{w}+1}....\sigma_{r_{b}-1} (-\sigma_{r_{b}}) 
\sigma_{r_{b}+1}... \sigma_{N}}.
\label{Acondition}
\end{eqnarray}
If and only if Eq.(\ref{Acondition}) is satisfied does the state
of Eq.(\ref{0inSz}) satisfy Eq.(\ref{spinfermivac}). 
Eq.(\ref{Acondition}) is equivalent to the demand that
\begin{eqnarray}
\sigma^{x}_{r_{w}} \sigma^{x}_{r_{b}} | 0 \rangle
= - \left( \sigma^{z}_{r_{w}+1} ... \sigma^{z}_{r_{b}-1} \right) | 0 \rangle.
\label{opcond}
\end{eqnarray}
In turn, Eq.(\ref{opcond}) is equivalent to the condition 
\begin{eqnarray}
| 0 \rangle = \mathcal{B}_{r_{w}} | 0 \rangle,
\label{brw1}
\end{eqnarray}
for all sites $\vec{r}_{w}$
with the operator definition of Eq.(\ref{brw}). 
Eq.(\ref{brw1}) along with the condition
of no vortices as dictated by reflection positivity \cite{Kitaev2006}
($I_h=1$) or an easier immediate direct inspection
of Eq.(\ref{EQ-fermion-model}) leads to the 
general solution for the Fermi vacuum state
in the spin basis (Eq.(\ref{fermion_vac})).
It is noteworthy that in the particular limit
$J_{z}>0$ with  $J_{x}=J_{y}=0$,any of the $2^{N/2}$ states
having $\sigma_{r_{w}}^{z} \sigma_{r_{b}}^{z} =1$ is
a ground state. Here, $N$ is the number of sites
and $(N/2)$ is the number of vertical ($\sigma^{z} \sigma^{z}$) 
bonds. As seen from our fermionization (e.g. 
Eq.(\ref{EQ-fermion-model})), this corresponds 
to the condition $\alpha_{r} (2n_{r}-1) =-1$ for
each vertical bond $r$ with $n_{r}= {d}_{r}^{\dagger} d_{r}$ the fermionic 
occupancy. In the sector $\alpha_{r}=1$ for all $r$,
the remaining ground state is that of the fermionic
vacuum derived above. As shown in Section(\ref{section-symmetry}),
the inversion of $\alpha_{r}$ on all sites $r$ of a given row  
leaves the spectrum invariant. If there are $L$ horizontal rows then
there are $2^{L}$ fermionic sectors that share the same spectrum.

\section{String correlators in matter Coupled Gauge theories}
\label{app2}

Here, we briefly review, 
the well known local symmetries of lattice gauge theories 
in order to clarify their similarity and the similarity
of the string correlators that they mandate to the 
string correlators in Kitaev's model. 
In Section(\ref{section-correlators}), 
we invoke precisely this analogy.
The crux of the selection rule
on the allowed correlations 
(which forces all correlators
be string type operators in both
gauge theories and Kitaev's
model) is
Elitzur's theorem. Elitzur's
theorem states that any
quantity which does not transform
as a singlet
under local (gauge) symmetries
must have a vanishing expectation 
value.

We now review 
Elitzur's theorem \cite{Elitzur} in its more
prominent use- that of gauge
theories. 
In theories of matter at lattice sites ($\sigma_{p}$) which are coupled 
to gauge fields ($U_{pq}$) which reside on links of the lattice, 
\cite{kogut} \cite{FS} the action is a sum of (i) 
a plaquette product of the gauge fields and (ii) a minimal coupling between matter fields and the gauge fields. To illustrate, consider the
simplest gauge theory 
[the Ising gauge theory] in which Ising gauge fields
are coupled to Ising matter fields. Here, the action
\begin{eqnarray}
S = - K \sum_{\Box} U_{pq}U_{ql}U_{ln}U_{np}
- J \sum_{pq} \sigma_{p} U_{pq} \sigma_{q},
\label{sg}
\end{eqnarray}
with $U_{pq} = \pm 1$ and $\sigma_{p} = \pm 1$.
Many early results were found 
by \cite{FS}. This action is invariant under
the local (gauge) transformations
$\sigma_{p} \to \eta_{p} \sigma_{p}, U_{pq} \to \eta_{p} U_{pq} 
\eta_{q}$ with, at any site $p$, 
$\eta_{p} = \pm 1$. Let us define gauge invariant link variables
by $z_{pq} \equiv \sigma_{p} U_{pq} \sigma_{q}$. The action is
a functional of $\{z_{pq}\}$. Any 
correlator which involves a product of any number of $z$'s
is invariant under the local 
gauge transformations and consequently does need not 
vanish by Elitzur's theorem \cite{Elitzur}. In matter coupled
gauge theories with $J \neq 0$ in Eq.(\ref{sg}), 
any correlator of the form 
\begin{eqnarray}
\langle \prod_{pq \in C} z_{pq} \rangle
\label{wil}
\end{eqnarray}
for any contour $C$ (either open or closed
or a contour $C$ which is the 
union of smaller open/closed contours) 
is gauge invariant and need not vanish
by Elitzur's theorem. For closed
contours $C$, the average of 
Eq.(\ref{wil}) is the ``Wilson loop'' \cite{kogut}
which makes an appearance also in ``pure'' ($J=0$)
gauge theories. These {\em string} type 
correlators (as well as related topological percolation 
transitions and crossovers \cite{percolation}) provide 
the only means for probing the behavior of this system. 
Precisely such {\em string} operators (both closed and 
open ended variants) are the sole non-zero
correlators than are allowed by symmetry 
in Kitaev's model [see Section(\ref{hior})].



\begin{thebibliography}{99}
\bibitem{fusion}
Generally, in a fusion of a general
individual tori of genus g=1 each having $N_{a}$ sites into a general
torus of genus g, the total number of spins is

\begin{eqnarray}
N_{s} = \sum_{a=1}^{g} N_{a} - \sum_{i} (n_{i}-1).
\end{eqnarray}

Here, $n_{i}$ is the number of tori to which a given site
$i$ is common. The second term above
removes the multiple counting of a given
site if it appeared in several tori prior
to the fusion. The total number $N_{s}$
of $S=1/2$ spins can be either
even or odd. The 13 site geometry
discussed earlier is a particular
instance of this with $N_{1,2,3} = 6$
(three hexagons) which has three
sites common to two tori and one point
common to all three tori. There,
we obviously have $N_{s} = 3 \times 
6 - 3 - 2 = 13$ (odd).
\end{thebibliography}
\end{document}